\documentclass{article}

\usepackage[english]{babel}

\usepackage[letterpaper,top=2cm,bottom=2cm,left=3cm,right=3cm,marginparwidth=1.75cm]{geometry}
\usepackage[T1]{fontenc}
\usepackage[latin1]{inputenc}
\usepackage{graphicx}
\usepackage{amsfonts, amsmath}
\usepackage{amssymb,mathtools}
\usepackage{amstext}
\usepackage{color}
\usepackage{dsfont}
\usepackage{subfig}

\begin{document}
\graphicspath{{figs/}}

\title{Importance of near- and far-field coupling for the synchronization of organ pipes}

\author{Jakub Sawicki$^{1,2}$}
\date{
    $^1$Potsdam Institute for Climate Impact Research\\%
    $^2$University of Applied Sciences Northwestern Switzerland\\[2ex]%
    \today
}

\maketitle

\begin{abstract} 
The synchronization of coupled organ pipes represents a nonlinear phenomenon with significant implications for both musical acoustics and nonlinear dynamics. This study investigates the coupling mechanisms governing synchronization, employing a theoretical framework based on delay-coupled Van der Pol oscillators. The oscillatory behavior of individual organ pipes is modeled as a self-sustained system driven by an air jet interacting with the pipe resonator. By incorporating experimental data and numerical simulations, we analyze the impact of finite separation between coupled pipes, introducing time-delayed interactions in the coupling function. Our findings reveal that synchronization can enhance pitch stability but may also induce sound attenuation due to destructive interference, particularly in prospect pipes. A key focus of this study is the decomposition of the coupling mechanism into near-field and far-field components, each contributing differently to the overall synchronization dynamics. Through bifurcation analysis and Arnold tongue characterization, we demonstrate how distance-dependent interactions influence the stability and frequency locking of coupled oscillators. The Fast Fourier Transform (FFT) is employed to extract spectral features from numerical data, ensuring precise comparison with experiments. Our results show strong qualitative agreement between modeled and observed synchronization effects, highlighting the importance of nonlinear coupling in organ pipe acoustics. This investigation advances the understanding of mutual interactions in coupled acoustic oscillators and provides a refined theoretical basis for predicting synchronization phenomena in complex aerodynamically driven systems.
\end{abstract}

\section{Introduction}
The physics of organ pipes constitutes a multidisciplinary research field, integrating concepts from nonlinear dynamical system theory \cite{FAB00,BAD13,FLU13}, aeroacoustic modeling \cite{HOW03}, and synchronization theory \cite{PIK01}. The organ, often referred to as the ``queen of instruments'', is distinguished by its grandeur and complex acoustic characteristics. In this study, we examine a nonlinear phenomenon in organ pipes: the synchronization of adjacent pipes. Recent investigations have employed both experimental and theoretical approaches \cite{ABE06,ABE09,FIS14,FIS16,SAW18a,SAW20}. From a musical standpoint, synchronization can have both advantageous and disadvantageous effects; it can stabilize the pitch of certain organ pipes, yet it may also lead to sound attenuation due to destructive interference, a phenomenon particularly observed in the prospect pipes of an organ \cite{RAY82,FLE78,STA01}.\\

A theoretical framework for understanding nonlinear interactions in organ pipes can be established by modeling a single pipe as a self-sustained oscillator. The oscillatory component in this system is the air jet, or ``air sheet'', emerging from the pipe mouth, while the resonator is the pipe body. Sound waves generated at the labium, the sharp edge at the upper opening of the pipe, propagate within the pipe and reinforce the oscillatory motion of the air sheet. The energy necessary for sustained oscillations is provided by a pressure reservoir beneath the pipe, ensuring a nearly constant flow rate. A schematic representation of an organ pipe is depicted in the right panel of Fig.\,\ref{goal}. Experimental and numerical investigations by Abel et al. \cite{ABE09,FIS16} have demonstrated that the dynamical behavior of an organ pipe can be effectively approximated by a Van der Pol oscillator.\\

Fischer \cite{FIS14,FIS16} further investigated the role of nonlinearities in sound generation and their influence on synchronization. In this study, we extend this analysis by examining the effect of finite separation between two coupled organ pipes, introducing a time delay into the coupling function. Specifically, we analyze bifurcation scenarios in a system of two delay-coupled Van der Pol oscillators, drawing on experimental configurations described by Bergweiler et al. \cite{BER06}.\\

We examine the synchronization phenomena of coupled organ pipes, where reflection and interaction effects can result in unintended sound attenuation. Experimental data indicate that these interactions are inherently nonlinear and complex. However, we demonstrate that a system of two delay-coupled Van der Pol oscillators serves as an effective model for capturing the primary dynamical features. By incorporating distance-dependent (or equivalently, time-delayed) coupling, we analytically investigate synchronization frequencies and bifurcation scenarios occurring at the Arnold tongue boundaries. A comparison of our theoretical predictions with experimental observations reveals qualitative agreement, particularly regarding the nonmonotonic modulation of the Arnold tongue shape, in contrast to the linear Arnold tongues observed in simpler coupled systems, such as the pipe-loudspeaker setup.\\

To further refine our analysis, we employ the Fast Fourier Transform (FFT), a fundamental algorithm in signal processing, numerical analysis, and various scientific and engineering applications. The FFT efficiently computes the Discrete Fourier Transform (DFT) and its inverse, facilitating the transformation of discrete signals between the time and frequency domains. By significantly reducing the computational complexity of the DFT, the FFT enables real-time signal analysis, image processing, and spectral analysis. Utilizing this algorithm, we achieve remarkable agreement between experimental and simulation results.\\

To enhance our understanding of synchronization in organ pipes, we compare experimental observations with a theoretical model based on fundamental principles of fluid mechanics and aeroacoustics. The Arnold tongue, a mathematical construct in parameter space describing the synchronization properties of coupled oscillators, provides a quantitative framework for this analysis. The experimentally determined Arnold tongue exhibits a nonlinear structure, reflecting the intricate coupling between interacting pipes. By developing and numerically integrating a coarse-grained model of two nonlinear coupled self-sustained oscillators \cite{FIS16}, we achieve a high degree of agreement with experimental data, particularly for separation distances in different field regimes. These methods contribute to a deeper comprehension of fundamental sound generation processes and the coupling mechanisms governing mutual interactions in acoustic oscillators. In this study, we examine the precise effects of the coupling mechanism, incorporating both near-field and far-field components.\\

In Sec.\,\ref{sec:model}, we introduce a mathematical framework based on two delay-coupled Van der Pol oscillators as a simplified model for coupled organ pipes. Section~\ref{sec:analytics} presents analytical methodologies aimed at providing deeper insights into the synchronization phenomena. The significance of selecting an appropriate coupling function is analyzed in Sec.\,\ref{sec:delay}. These findings are then compared with experimental acoustic data in Sec.\,\ref{sec:comparison}. Finally, we summarize our conclusions in Sec.\,\ref{sec:conclusion}.

\section{A model of coupled organ pipes}
\label{sec:model}

To obtain a deeper insight into the synchronization phenomena of two coupled organ pipes we model the pipes by Van der Pol oscillators with delayed cross-coupling \cite{SAW18a}:
\begin{equation}
\label{ausgang}
\ddot{x}_i+{\omega_i}^2x_i-{\mu}\left[\dot{x}_i-\dot{f}(x_i)+{\kappa}(\tau)x_j(t-\tau)\right]=0,
\end{equation}
where $i,j=1,2$. These equations represent a harmonic oscillator with an intrinsic angular frequency $\omega_i$, supplemented with linear and nonlinear damping of strength $\mu > 0$. The nonlinear damping can be described by the nonlinear function
\begin{eqnarray}
\label{ausgang11}
f(x_i)=\frac{{\gamma}}{3}x_i^3,
\end{eqnarray}
where $\gamma$ is the anisochronicity parameter and $\dot{f}(x_i)={\gamma} x_i^2 \dot{x_i}$. The coupling delay is $\tau$, and the delay-dependent coupling strength in Eq.\,\eqref{ausgang} is $\kappa(\tau)$. For the clarity of the calculations we keep the coupling strength constant in the first part of the paper ($\kappa(\tau)=\kappa$), but all analytical results hold as well for general $\kappa(\tau)$. Since for synchronization the frequency difference of the two oscillators is important, we introduce the detuning parameter $\Delta \in \mathbb{R}$ by

\begin{eqnarray}
\label{ausgang12}
\omega_1^2=\omega_2^2+\mu \Delta. 
\end{eqnarray}

\begin{figure}
\centering
\includegraphics[width=.5\linewidth]{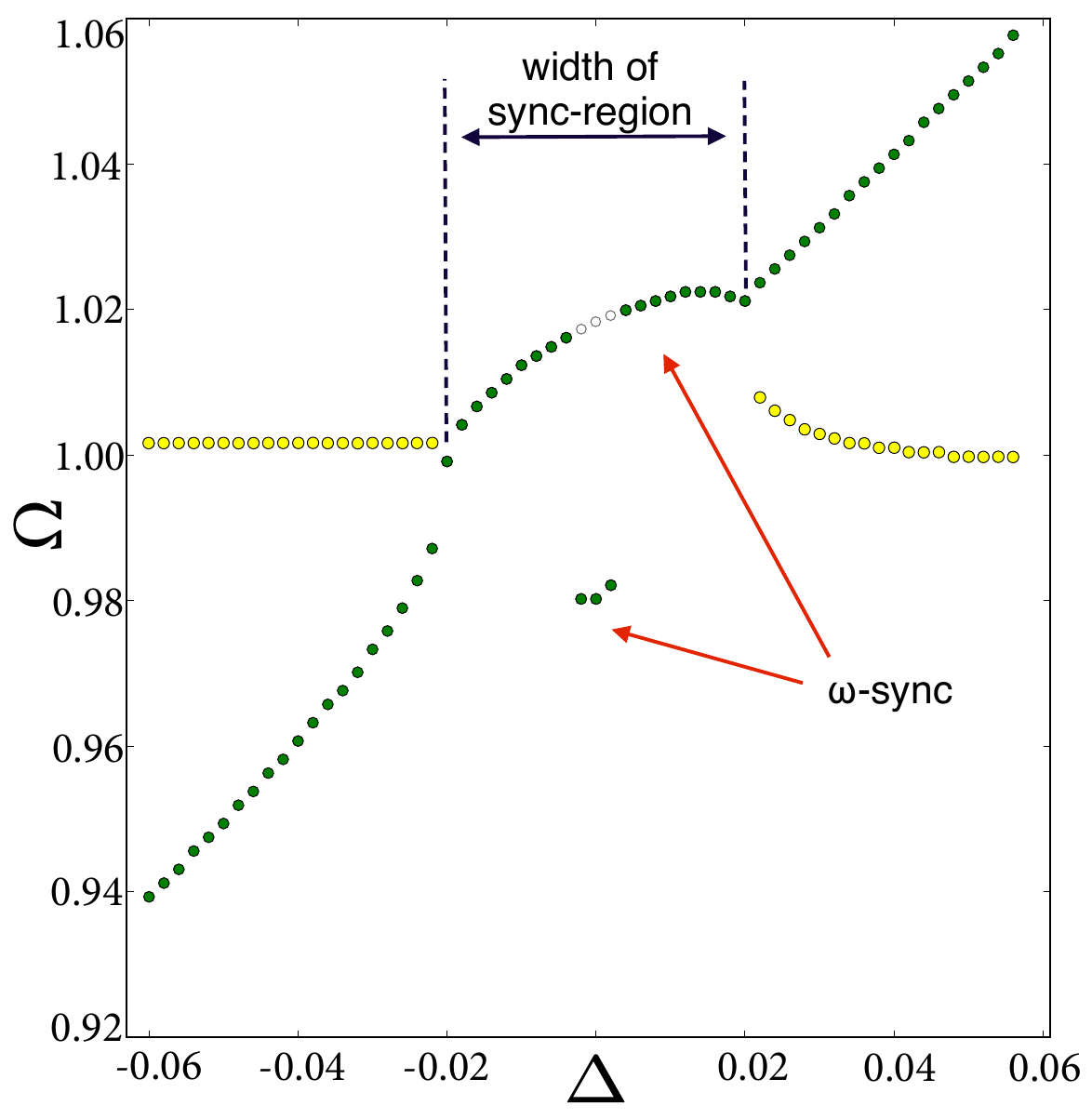}
\includegraphics[width=.35\linewidth]{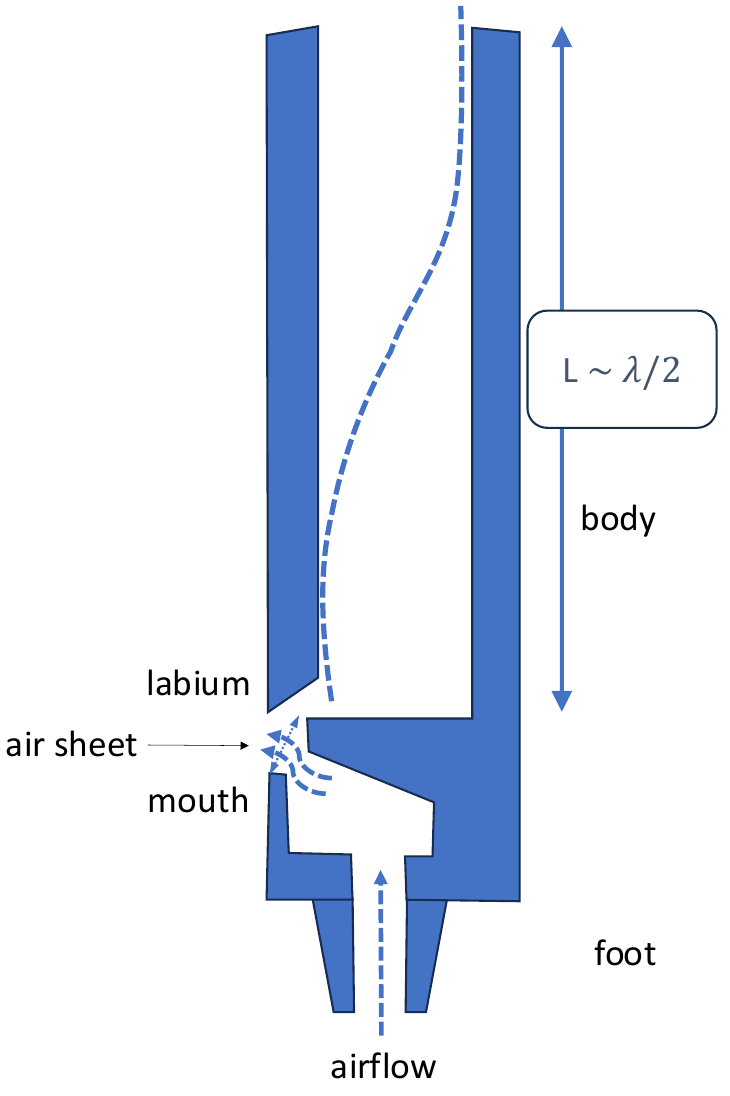}
\caption{\label{goal}\raggedright Synchronization of organ pipes: (left) Angular frequency $\Omega$ of oscillator $x_1$ (dark green circles), and oscillator $x_2$ (light yellow circles) versus the detuning $\Delta$ of the oscillators. Full (empty) circles correspond to symmetric (non-symmetric) initial conditions. Parameters:  $\omega_{2}=1$, $\mu = 0.1$, $\gamma=1$, $\kappa=0.4$, $\tau=0.1\pi$. Figure taken from \cite{SAW18a}. (right) Organ pipe as a self-sustained oscillator: The pressure waves driven by the supplied airflow trigger the ``air sheet'' which exits at the pipe mouth. The open organ pipe has a length of $L\sim \lambda/2$, where $\lambda$ denotes the wavelength of the tone of the pipe.}
\end{figure}

Figure~\ref{goal} illustrates the frequency locking behavior obtained through numerical simulations of Eq.\,\eqref{ausgang} under symmetric initial conditions (left panel). The plot depicts the angular frequencies $\Omega$ as a function of the detuning $\Delta$ between two delay-coupled Van der Pol oscillators. A distinct synchronization region is observed, characterized by a sharp transition to frequency locking. Within this region, the in-phase synchronized solution (lower branch) occurs only for small values of $|\Delta|$, while for larger detuning, the system transitions to the anti-phase synchronized state (upper branch), even when starting from symmetric initial conditions (represented by full circles). Notably, for small $|\Delta|$, the anti-phase solution also emerges when non-symmetric initial conditions are applied (empty circles). This study aims to analyze key characteristics of synchronization, including the synchronization frequency, the width of the synchronization region, the phase difference in the synchronized state, its stability, and the bifurcation scenarios at the synchronization boundaries.

\section{Analytical approaches and Arnold tongue}
\label{sec:analytics}

The method of averaging (quasiharmonic reduction) describes weakly nonlinear oscillations in terms of slowly varying amplitude and phase. For $ {\mu}= 0$ the uncoupled system reduces to the harmonic oscillator $\ddot{x}_i+{\omega_i}^2x_i=0$ with solution
\begin{eqnarray}
\label{hl}
{x_i}=R_i\sin({\omega_i}t+\phi_i), 
\end{eqnarray}
with constant amplitude $R_i$ and phase $\phi_i$. 
For $0<{\mu}\ll 1$ we look for a solution in the form Eq.\,(\ref{hl}) but assume that the amplitude $R_i \ge 0$ and the phase $\phi_i$ are time-dependent functions:

\begin{eqnarray}
\begin{aligned}
\label{fhl1}
{x_i}&=R_i(t)\sin({\omega_i}t+\phi_i(t)),\\
\dot{x}_i&=R_i(t){\omega_i}\cos({\omega_i}t+\phi_i(t)),
\end{aligned}
\end{eqnarray}

where terms involving the slowly varying functions $\dot{R}_i$, $\dot{\Phi}_i$ are neglected. Without loss of generality, we choose ${\omega_{2}} = 1$. For small $\mu$ we use the method of averaging, assuming that the product $\mu\tau$ is small, and Taylor expand $R_i(t-\tau)$ and $\phi_i(t-\tau)$ in the following way:
\begin{eqnarray}
\label{taylor1}
R_i(t-\tau)=R_i(t)-\tau\dot{R}_i(t)+\frac{\tau^2}{2}\ddot{R}_i(t)+\dots\quad. 
\end{eqnarray}
We introduce the phase difference $\psi(t)=\phi_1(t)-\phi_2(t)$. Defining a new time scale $\tilde{t}=\frac{2t}{\mu}$, we find the equations which describe the system \eqref{ausgang} on a slow time scale:
\begin{eqnarray}
\label{R1}
\dot{R}_{1/2}(\tilde{t})={R_{1/2}(\tilde{t})}\left(1-\frac{{\gamma}{R_{1/2}(\tilde{t})}^2}{4}\right) \mp{\kappa}R_{2/1}(\tilde{t})\sin(\psi(\tilde{t})+\tau), 
\end{eqnarray}
\begin{eqnarray}
\label{Psi}
\dot{\psi}(\tilde{t})=-\Delta+{\kappa}\left[\frac{R_1(\tilde{t})}{R_2(\tilde{t})}\cos(\psi(\tilde{t})-\tau)-\frac{R_2(\tilde{t})}{R_1(\tilde{t})}\cos(\psi(\tilde{t})+\tau)\right].
\end{eqnarray}

For the sake of simplicity, we omit the tilda\,$\sim$ in the following. The combined effect of the method of averaging and the truncation of the Taylor expansion in $\tau$, is a reduction of the infinite-dimensional problem to a finite-dimensional one (valid only if the product $\mu\tau$ is small). This key step enables us to handle the original delay differential equation as a system of ordinary differential equations \cite{WIR02, SEM15}. We now have two dynamical equations \eqref{R1} for the amplitudes $R_1$ and $R_2$, and one equation \eqref{Psi} for the phase difference $\psi(t)$, which is also called the {\em slow phase}. The latter equation is a generalized Adler equation \cite{ADL73} and contains the main features of synchronization.

The equilibria of the Adler equation correspond to the locking of phase and frequency, since the difference between the phases is constant. To investigate the stability and bifurcation scenario of such fixed points we take a closer look at the generalized Adler equation~\eqref{Psi}, written in general form:
\begin{equation}
\label{adler}
\begin{split}
\dot{\psi}(t)=-\Delta+{\kappa}q({\psi}(t)),
\end{split}
\end{equation}
where the averaged forcing term $q({\psi}(t))$ is the $2\pi$-periodic function
\begin{equation}
\label{q}
\begin{split}
q({\psi}(t))=\tfrac{R_1(t)}{R_2(t)}\cos\left[\psi(t)-\tau\right]-\tfrac{R_2(t)}{R_1(t)}\cos\left[\psi(t)+\tau\right].
\end{split}
\end{equation}
The generalized Adler equation~\eqref{Psi} is a valuable tool for the calculation of the Arnold tongue, which is one of the main characteristics of synchronization in nonlinear systems. 

Furthermore, we gain information about the stability of the synchronization state: For $\psi=0$ (or, equivalently, $\psi=2\pi$) we have an unstable equilibrium, and for $\psi=\pi$ (anti-phase oscillation) a stable equilibrium, since $\dot{\psi}<0$ for $\psi>0$ and $\dot{\psi}>0$ for $\psi<0$. This is in accordance with experimental results \cite{FIS14}, as discussed in Fig.\,\ref{realpipeex}a below. The experimentally observed decrease of the amplitude at $\Delta = 0$ indicates an anti-phase oscillation \cite{ABE06}. The in-phase and anti-phase mode correspond to the enhancement or cancellation of sound in organ pipe experiments, respectively. As we recognize from Fig.\,\ref{goal} the center of the synchronization region plays a special role. This motivates a first investigation of the solutions and their stability for vanishing detuning $\Delta=0$.

In Fig.\,\ref{goal}, the upper frequency branch in the synchronization region stays in the stable anti-phase mode for non-zero detuning $\Delta$ (in congruence with experimental data \cite{ABE06}), whereas the lower branch, i.e., the stable in-phase mode, which is close to its instability point, is only observable in a small range of $\Delta$. 


\begin{figure}
\centering
\includegraphics[width=.45\linewidth]{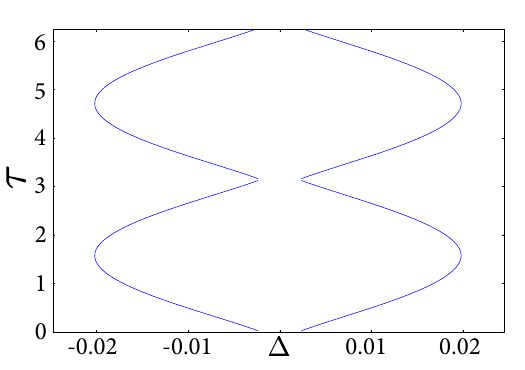}
\caption{\label{arnoldtau}\raggedright The synchronization region in the parameter plane of delay time $\tau$ and frequency detuning $\Delta$. The plot shows an analytically computed Arnold tongue for the coupling strength $\kappa=0.2$, $\omega_2=1$, $\mu = 0.1$, $\gamma=1$. One can see the symmetric, $\pi$-periodic boundaries of the tongue. Figure taken from \cite{SAW18a}.}
\end{figure}

A key theoretical aspect of synchronization is the transition to frequency locking, typically analyzed in the parameter space of frequency detuning $\Delta$ and coupling strength $\kappa$. In systems with delayed coupling, the time delay $\tau$ plays a role comparable to the coupling strength, significantly influencing the synchronization behavior. The synchronization region in the $(\kappa, \Delta)$ or $(\tau, \Delta)$ parameter plane is commonly referred to as the Arnold tongue, a fundamental characteristic of synchronizing nonlinear systems.

To characterize this behavior, we analytically determine the synchronization region in the $(\tau, \Delta)$ plane, applying the methods outlined in the previous section. The results, shown in Fig.\,\ref{arnoldtau}, exhibit excellent agreement with numerical simulations of Eq.\,(\ref{ausgang}). A notable feature of the Arnold tongue boundaries is their periodic modulation with a period of $\pi$ as $\tau$ varies. This periodicity arises because, for our choice of $\omega_2 = 1$, the natural period of the uncoupled harmonic oscillator is $2\pi$.

Furthermore, our analysis remains valid when the constant coupling strength $\kappa$ is replaced by a time-delay-dependent coupling function $\kappa(\tau)$, as discussed in Sec.\,\ref{sec:delay}. This extension provides a more accurate representation of experimental observations and highlights the critical role of delay-dependent interactions in synchronization dynamics.


Equations (\ref{R1}) and (\ref{Psi}) possess two equilibrium solutions, an in-phase and an anti-phase mode, as has been demonstrated \cite{SAW18a}. The in-phase and anti-phase mode correspond to the enhancement or cancellation of sound in organ pipe experiments, respectively. As we recognize from Fig.\,\ref{goal} the center of the synchronization region plays a special role. This motivates a first investigation of the solutions and their stability for vanishing detuning $\Delta=0$. Such a special parameter setting reduces the technical difficulties and nevertheless allows us to make a qualitative and qualitative analysis of our problem. We find the following equilibrium points for Eqs.\,(\ref{R1}) and (\ref{Psi}) which in turn correspond to the in-phase and anti-phase mode of Eq.\,\eqref{ausgang}:

\begin{equation}
\label{loes1}
\begin{split}
\psi_{in}=0 \Leftrightarrow R_1= R_2=\frac{2}{\sqrt{\gamma}}\sqrt{1-{\kappa}\sin \tau},\\
\psi_{anti}=\pi \Leftrightarrow R_1= R_2=\frac{2}{\sqrt{\gamma}}\sqrt{1+{\kappa}\sin \tau}.
\end{split}
\end{equation}
In order to determine the stability of the in-phase and anti-phase mode, we linearize Eqs.\,(\ref{R1}),(\ref{Psi}) around the equilibrium points, which gives the Jacobian matrix $J$ of the system\,(\ref{R1}),(\ref{Psi}):

\begin{equation}
\label{Jey}
J=\left(
\begin{array}{ccc}
 1-\frac{3}{4}  {\gamma} R_1^2 & -\kappa \sin(\psi+\tau) & -\kappa R_2 \cos(\psi+\tau) \\
 \kappa \sin(\psi-\tau) & 1-\frac{3}{4}  {\gamma} R_2^2 & \kappa R_1 \cos(\psi-\tau) \\
 \kappa \left[\frac{R_2 \cos(\psi+\tau)}{R_1^2}+\frac{\cos(\psi-\tau)}{R_2}\right] & -\kappa \left[\frac{\cos(\psi+\tau)}{R_1}+\frac{R_1 \cos(\psi-\tau)}{R_2^2}\right] & \kappa \left[\frac{R_2 \sin(\psi+\tau)}{R_1}-\frac{ R_1 \sin(\psi-\tau)}{R_2}\right]\\
\end{array}
\right).
\end{equation}

The eigenvalues $\lambda_{i}$, $i=1,2,3$ of the Jacobian matrix evaluated at these equilibrium points determine their linear stability. They are calculated from the characteristic equation
\begin{eqnarray}
\label{deteq}
\det(J-\lambda_{i}\mathsf{I}) = 0,
\end{eqnarray}
in dependence on the system parameters $\gamma$, $\kappa$, and $\tau$. The stability of the in-phase and anti-phase synchronization modes in delay-coupled Van der Pol oscillators is analyzed by deriving the characteristic equations for both cases \cite{SAW18a}. The stability boundaries, determined by saddle-node bifurcations, are mapped in the parameter space of coupling strength $\kappa$ and time delay $\tau$. For the in-phase mode, the stability regions are defined by bifurcation curves, which indicate transitions between stable and unstable regimes as $\tau$ varies. Similarly, for the anti-phase mode, corresponding bifurcation curves separate stable and unstable regions. 

The bifurcation structure reveals that stability is periodically modulated with $\tau$, exhibiting a periodicity of $\pi$. At $\tau = 0$, both synchronization modes are stable. However, increasing $\tau$ leads to alternating stability windows, where either the in-phase or the anti-phase mode becomes unstable depending on the bifurcation conditions. Notably, bistability occurs around $\tau = 0$ and $\tau = \pi$, where both modes can coexist. Numerical simulations confirm these theoretical predictions, demonstrating how varying $\kappa$ affects the stability transitions. For instance, at $\kappa = 0.4$, the in-phase mode loses stability in a certain range of $\tau$, while for a lower coupling strength ($\kappa = 0.2$), the in-phase mode remains unstable over a broader range, eliminating bistability. These findings highlight the intricate interplay between coupling strength and delay in determining synchronization stability, which is crucial for understanding delayed coupled oscillatory systems.

The describing function method (also called the method of harmonic balance) uses frequency domain techniques to investigate limit cycle behavior in nonlinear systems \cite{SAW20}. An approach with the describing function method determines the synchronization frequency $\omega_{\textsl{\textrm{s}}}$, if the phase difference $\psi$ is known. The method of averaging on the other hand yields the generalized Adler equation determining the dynamics of $\psi$. The describing function method determines the synchronization frequency $\omega_s$ given a known phase difference $\psi$. The method of averaging yields the generalized Adler equation~\eqref{adler}, governing the dynamics of $\psi$. Numerical analysis using the Adler equation reveals a stable equilibrium at $\psi=\pi$ and an unstable one at $\psi=0$.

\begin{figure}
\centering
\includegraphics[width=.6\linewidth]{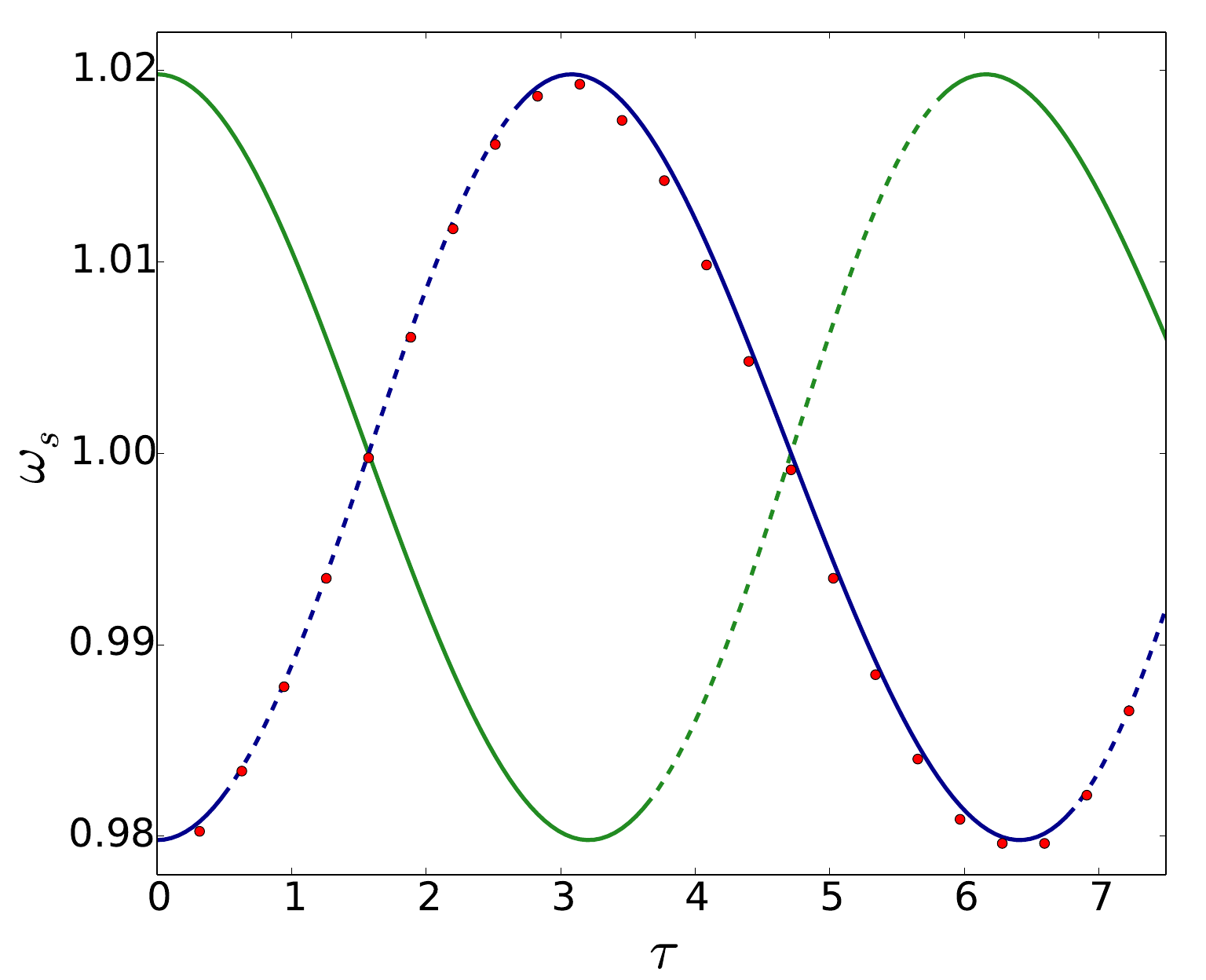}
\caption{\raggedright Comparison of the analytic (line) and numeric (dots) results of the synchronization frequency $\omega_{\textsl{\textrm{s}}}$ versus the delay time $\tau$ for $\Delta=0, \omega_1=\omega_2=1$, $\mu = 0.1$, $\gamma=1$, $\kappa=0.4$. The analytic solution gives an in-phase (dark blue line) and an anti-phase mode (light green line), while the numeric solution (dots) with symmetric initial conditions only reproduces the in-phase mode. Solid line means stable solution, whereas dashed line stands for an unstable one. Figure taken from \cite{SAW18a}.}
\label{comp1}
\end{figure}

In order to compare the numerical simulations with the results of the describing function method, we set $\omega_1=\omega_2=1$, i.e., $\Delta=0$. In this case the synchronization frequency can be simplified to 

\begin{equation}
\label{omega_simpl2}
\begin{split}
\omega_{\textsl{\textrm{s}}}^2=1 \pm\mu\kappa\cos(\omega_{\textsl{\textrm{s}}}\tau)
\end{split}
\end{equation}

where $+$ and $-$ correspond to anti-phase and in-phase oscillations, respectively. In Fig.\,\ref{comp1}, we plot the synchronization frequency $\omega_{\textsl{\textrm{s}}}$ versus the time delay $\tau$ for $\Delta=0$. The congruence between the numerical (from Eq.\,(\ref{ausgang})) and analytical result for the in-phase mode (dark blue line, from Eq.\,(\ref{omega_simpl2})) is excellent. It is remarkable that the synchronization frequency $\omega_{\textsl{\textrm{s}}}$ is modulated around the single oscillator frequencies $\omega_i=1$ in dependence upon the delay time. For small delay time, for instance, the in-phase oscillation frequency is lowered, while the anti-phase oscillation frequency (light green line) is increased. The stability of the two branches changes as $\tau$ is varied: At the extrema of the frequency curve in Fig.\,\ref{comp1} we find bistability.

Figure~\ref{comp1} presents the dependence of $\omega_s$ on $\tau$ for $\Delta=0$. The numerical results (dots, derived from Eq.\,\eqref{ausgang}) show excellent agreement with the analytical prediction for the in-phase mode (dark blue line, from Eq.\,\eqref{omega_simpl2}). Notably, $\omega_s$ oscillates around the natural frequency $\omega_i=1$ as a function of $\tau$. For small $\tau$, the in-phase frequency is reduced, while the anti-phase frequency (light green line) is increased. 

The stability of these branches varies with $\tau$. At the extrema of the frequency curve in Fig.\,\ref{comp1}, bistability is observed. For symmetric initial conditions, the in-phase mode emerges as the numerical solution across all delay times, even in regions where it is unstable. However, for non-symmetric initial conditions, the anti-phase mode would be found instead. In Fig.\,\ref{goal}, the upper frequency branch in the synchronization region remains in the stable anti-phase mode for $\Delta \neq 0$, consistent with experimental findings \cite{ABE06}. In contrast, the in-phase mode, being close to its instability threshold, is observable only within a limited range of $\Delta$.


\section{Dependence on near- and far-field coupling}
\label{sec:delay}

\begin{figure}
\centering
\includegraphics[width=.65\linewidth]{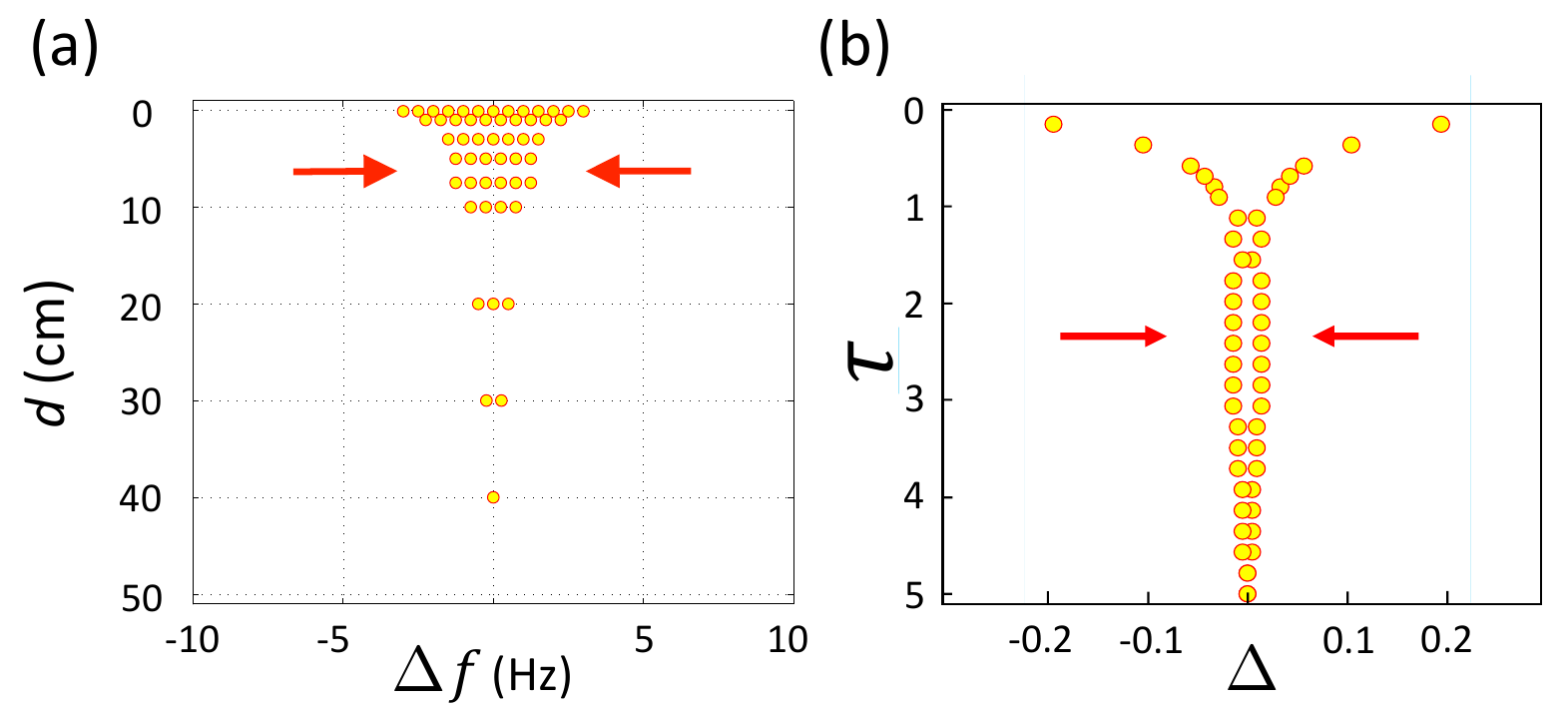}
\caption{\label{coupling_function}\raggedright Comparison of experiment and theory: Arnold tongue in the plane of coupling strength $\kappa$ vs. detuning $\Delta$: (a) experiment \cite{FIS14,FIS16}, where the coupling strength is given by the distance $d$ (cm) of the organ pipes, and (b) numerical result of Eq.\,\eqref{ausgang} with $\omega_2=1$, $\mu = 0.1$, $\gamma=1$, $\kappa_1=\kappa_n=\kappa_2=\kappa_f=0.04$. The red arrows in (a) and (b) indicate the non-monotonic behavior of the Arnold tongue, respectively.}
\end{figure}

In previous studies \cite{SAW18a}, the synchronization of organ pipes has been analyzed using a model of two delay-coupled Van der Pol oscillators with a constant coupling coefficient. In this work, we extend this approach by introducing a time-delay-dependent coupling strength, $C(\tau)$, replacing the constant coupling term. This refined model accounts for both near-field ($\sim \tfrac{1}{\tau^2}$) and far-field ($\sim \tfrac{1}{\tau}$) contributions. A time-dependent coupling strength is essential for accurately capturing the non-monotonic behavior of the Arnold tongue, particularly its pronounced widening at small delay times $\tau$, as observed experimentally (see Fig.\,\ref{coupling_function}a for $d=5\,\text{cm}$). Instead of a constant coupling coefficient $\kappa$ as in Eq.\,\eqref{ausgang}, we employ a coupling strength $\kappa(\tau)$ that explicitly depends on the time delay $\tau$.\\

The delay in coupling arises because the sound propagates over a finite distance $d$ between the pipes, and its strength is distance-dependent due to attenuation, which follows the radiation characteristics of a spherical wave emitted from the pipe mouth. The coupling strength $\kappa(\tau)$ incorporates two distinct terms: a near-field component ($\propto \frac{1}{\tau^2}$) and a far-field component ($\propto \frac{1}{\tau}$), each characterized by positive coefficients $\kappa_1 = \kappa_n$ and $\kappa_2 = \kappa_f$, respectively:

\begin{equation}
\label{eq:near_far}
\begin{split}
\kappa(\tau) = \frac{\kappa_1}{\tau^2} + \frac{\kappa_2}{\tau},
\end{split}
\end{equation}
where $\kappa_1=\kappa_n$ and $\kappa_2=\kappa_f$ are the near-field and far-field coupling parameters and $\tau$ is the time delay. 

\begin{figure}
\centering
\includegraphics[width=.75\linewidth]{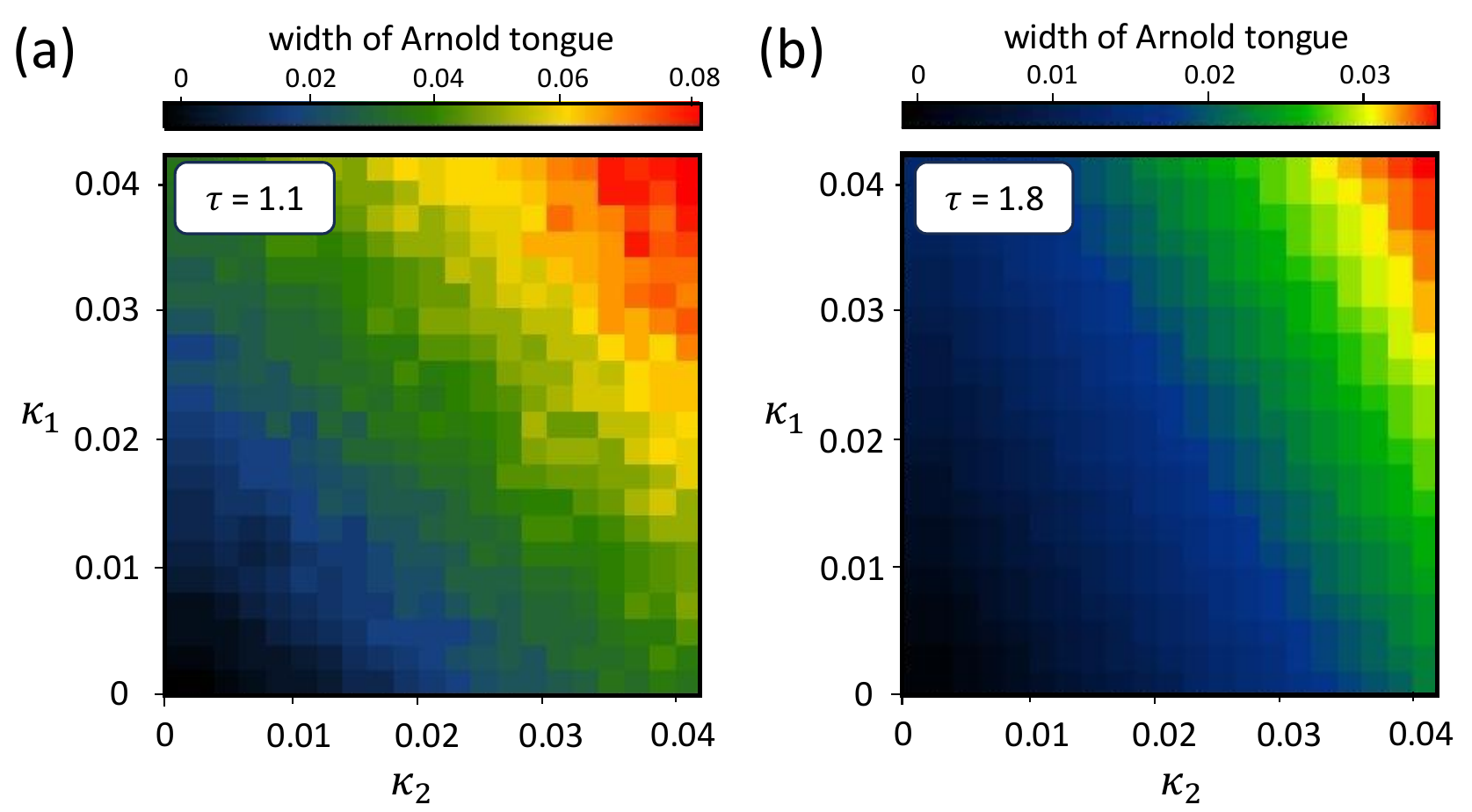}
\caption{\raggedright Dependence on near- and far-field parameters: Shown is the dependence of the with of the synchronization region of the Arnold tongue against $\kappa_1$ (near-field coupling) and $\kappa_2$ (far-field coupling). The width of the synchronization region is color-coded and varies comparing $\tau=1.1$ in (a) and $\tau=1.8$ in (b). The other parameters are given by $\omega_1 =1$, $\mu = 0.1$, $\gamma=1$.}
\label{near_far}
\end{figure}

By substituting $\kappa(\tau)$ from Eq.\,\eqref{eq:near_far} into the governing equation,

\begin{equation}
\label{ausgang_end}
\ddot{x}_i+{\omega_i}^2x_i-{\mu}\left[\dot{x}_i-\dot{f}(x_i)+{\kappa}(\tau)x_j(t-\tau)\right]=0,
\end{equation}

we analyze the boundaries of the Arnold tongue as a function of near- and far-field coupling contributions. The synchronization region's width varies with the delay time $\tau$, as depicted in Fig.\,\ref{near_far}. In Fig.\,\ref{near_far}a, where $\tau=1.1$, the Arnold tongue exhibits greater width, whereas increasing the delay time to $\tau=1.8$ in Fig.\,\ref{near_far}b results in more sharply defined synchronization regions. A fundamental distinction between these two delay times lies in the asymmetry of the resulting representations. For the smaller delay time $\tau=1.1$ in Fig.\,\ref{near_far}a, the coupling matrix remains symmetric, meaning that interchanging the near-field ($\kappa_1 = \kappa_n$) and far-field ($\kappa_2 = \kappa_f$) coupling coefficients does not affect the width of the Arnold tongue.\\

In contrast, for the larger delay time $\tau=1.8$ in Fig.\,\ref{near_far}b, the matrix exhibits asymmetry, implying that swapping the coupling parameters alters the width of the Arnold tongue. This observation suggests that the shape of the Arnold tongue can be precisely adjusted by modifying the relative contributions of near-field and far-field coupling. This indicates that the Arnold tongue shape can be fine-tuned by adjusting the relative contributions of near-field and far-field coupling. Notably, for small delay times, a significant widening of the Arnold tongue is observed, which is consistent with experimental findings (see Fig.\,\ref{coupling_function}b). This widening effect is attributed to the strong increase in coupling strength for small $\tau$, as described by Eq.\,\eqref{eq:near_far}:

\begin{equation}
\lim_{\tau \to 0}\kappa(\tau) = \infty.
\end{equation}

Another key consequence of the delay-dependent coupling strength is the destabilization of the in-phase solution for $\tau=0$ \cite{SAW18a}. Specifically, in the case where $\kappa_n = \kappa_f = \kappa$, the far-field coupling term $\kappa_2(0)$ satisfies:

\begin{equation}
\lim_{\tau \to 0}\kappa_2(\tau) = \frac{\tau^2}{(1+\tau)\sin\tau}= 0.
\end{equation}

As a result, for $\tau=0$, only the anti-phase mode remains stable. Introducing a delay-dependent coupling function $\kappa(\tau)$ thus leads to a distortion of the bifurcation curves, altering the stability properties of the synchronization modes.

\section{Comparison with acoustic experiments} 
\label{sec:comparison}

A direct comparison between a complex physical experiment and a simplified oscillator model presents significant challenges. An organ pipe, as a physical system, exhibits a rich spectrum of overtones, nonlinear interactions, and intricate aeroacoustic behavior influenced by factors such as turbulence, boundary layer effects, and acoustic feedback. In contrast, mathematical models often employ simplifications to capture only the most essential dynamical features. One such model is the Van der Pol oscillator, which serves as a fundamental representation of self-sustained oscillations in nonlinear systems. Despite its simplicity, the Van der Pol oscillator can exhibit a wide range of dynamical behaviors, including limit cycles, synchronization, and bifurcations, making it an effective tool for studying coupled oscillatory systems.\\

The value of employing such a minimalistic model lies in its ability to reveal fundamental principles governing the underlying physics. In dynamical systems theory, it is widely accepted that complex macroscopic behavior can emerge from relatively simple low-dimensional models. In this context, our analysis demonstrates that key aspects of the synchronization phenomena observed in organ pipes can be qualitatively and quantitatively reproduced using a simple oscillator model. This approach not only enhances our understanding of the fundamental mechanisms at play but also provides an analytical framework that can be generalized to a broader class of coupled oscillatory systems beyond acoustics, including biological, mechanical, and electronic oscillators.\\

To analyze the frequency content of both experimental and numerical data, we employ the Fast Fourier Transform (FFT). The FFT is a cornerstone algorithm in signal processing, numerical analysis, and various engineering and scientific disciplines. It efficiently computes the Discrete Fourier Transform (DFT) and its inverse, allowing for the transformation of discrete signals between the time and frequency domains. The direct computation of the DFT has a computational complexity of $O(N^2)$, making it impractical for large datasets. However, the FFT algorithm reduces this complexity to $O(N \log N)$, vastly improving efficiency and enabling real-time spectral analysis in various applications.\\

Since its introduction by Cooley and Tukey in 1965, the FFT has become an indispensable tool in numerous fields, including telecommunications, audio and speech processing, medical imaging, and computational physics. In our analysis, the FFT is employed to extract dominant frequency components from both experimental recordings and numerical simulations. This facilitates a direct comparison of spectral characteristics, enabling the identification of synchronization regimes and transitions between different oscillatory states. Additionally, by analyzing higher harmonics and spectral broadening, we can gain further insights into nonlinear effects and coupling mechanisms present in the system.\\

The synchronization scenarios obtained from our model exhibit strong qualitative agreement with experimental observations. This is particularly evident when comparing the characteristic features of synchronization regions derived from numerical simulations with those observed in experimental data. Despite the inherent simplifications in our model, it successfully captures essential aspects of the nonlinear coupling between oscillators, providing valuable insights into the underlying dynamics.\\

\begin{figure}
\centering
\includegraphics[width=.7\linewidth]{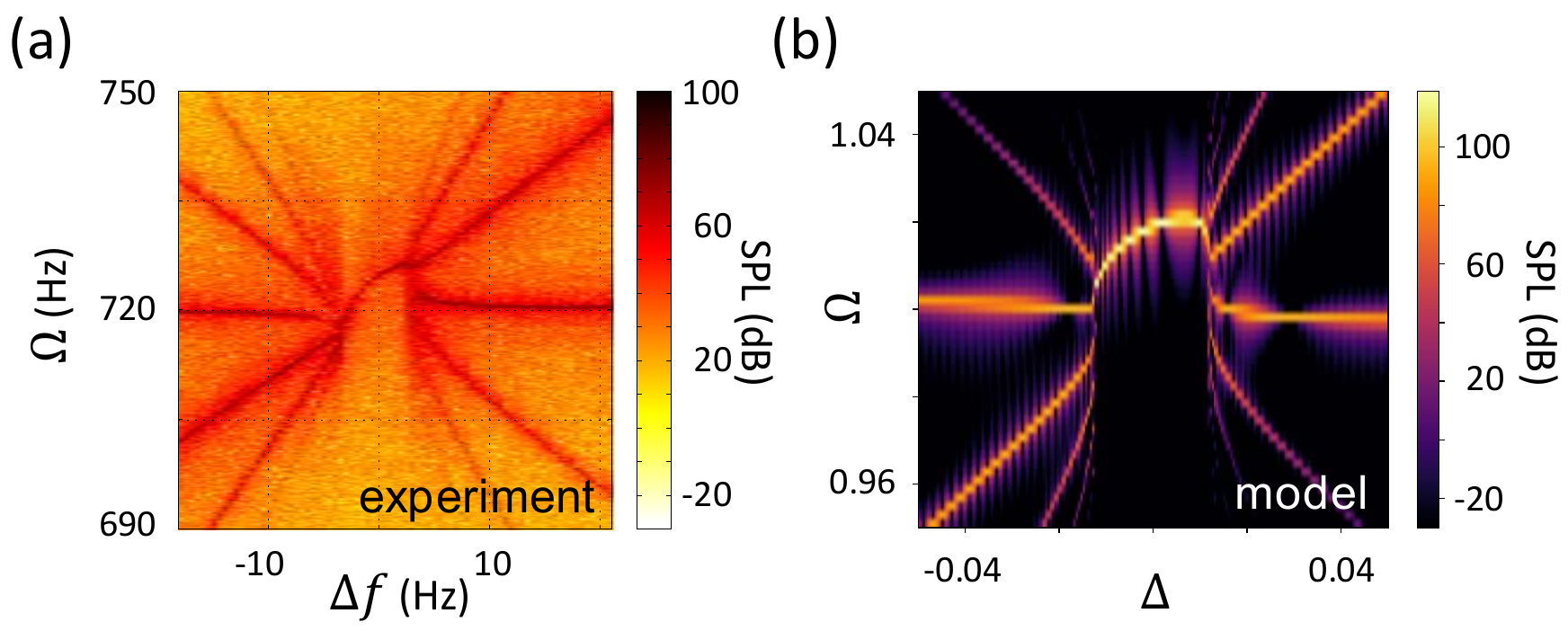}
\caption{\label{realpipeex}\raggedright Comparison of synchronization regions in experiment and theory: (a) experimentally observed sound pressure level (SPL) in the plane of observed frequency $\Delta$ versus frequency detuning $\Delta f$ (in Hz) \cite{FIS14}, and (b) numerically computed SPL using the Fast Fourier Transform (FFT) in the plane of angular frequency $\Omega$ versus dimensionless detuning $\Delta$, with parameters $\omega_2=1$, $\mu = 0.1$, $\gamma=1$, $\kappa=0.4$, $\tau=1.1\pi$.}
\end{figure}

For a direct visual comparison between experimental and theoretical results, we present the synchronization region as a function of frequency detuning in Fig.\,\ref{realpipeex}. Both plots display strikingly similar features, particularly in the structure of the transition regions at the boundaries of the locking interval and the concave curvature of the synchronization region. These similarities suggest that the primary mechanisms governing synchronization in coupled organ pipes can be effectively described using a low-dimensional model. Moreover, this reinforces the robustness of theoretical approaches based on coupled oscillator models in predicting real-world nonlinear phenomena.\\

Another important aspect of our analysis is the role of initial conditions in determining synchronization modes. As observed in our numerical simulations, different initial conditions can lead to distinct synchronization states, even for the same parameter values. This highlights the presence of multistability, where multiple stable solutions coexist within a certain parameter regime. The presence of multistability is a well-documented feature in nonlinear systems and has been observed in various physical, biological, and engineering contexts. In the context of coupled organ pipes, this suggests that small variations in experimental setup, such as the initial phase difference between oscillators or minor perturbations in system parameters, can influence the observed synchronization state.\\

Furthermore, in Fig.\,\ref{coupling_function} of Sec.\,\ref{sec:delay}, we have compared the experimentally observed Arnold tongue with its analytically calculated counterpart in the plane of coupling strength $\kappa$ and frequency detuning $\Delta$. The Arnold tongue describes the region in parameter space where stable synchronization occurs, with its characteristic wedge-shaped structure determined by the interplay between coupling strength and detuning. In experiments, the effective coupling strength is controlled by the spatial separation between two organ pipes, which influences the strength of acoustic interaction.\\

Our results reveal an Arnold tongue with nonlinear, curved boundaries -- a remarkable feature that manifests even for small delay times $\tau$ and aligns well with experimental data \cite{FIS14,FIS16}. The curvature of the Arnold tongue's boundaries suggests that nonlinearities in the coupling mechanism play a significant role in shaping the synchronization region. This can be further refined in our model by introducing a delay-dependent coupling function $\kappa(\tau)$, as discussed in Sec.\,\ref{sec:delay}. A more precise understanding of this dependence could provide further insights into the effects of time-delayed interactions and their influence on synchronization dynamics.


\section{Conclusion} 
\label{sec:conclusion}

In this study, we investigated the synchronization of organ pipes through the lens of nonlinear dynamics, focusing on the crucial role of near- and far-field coupling in shaping synchronization behavior. The finite spatial separation between pipes naturally introduces a coupling delay, which significantly affects their collective dynamics. To model this phenomenon, we employed two coupled Van der Pol oscillators interacting via a dissipative, direct, and delayed coupling. By systematically varying the coupling strength \(\kappa\) and delay time \(\tau\), we explored how different coupling regimes influence the synchronization properties of the system. \\

A key aspect of our analysis was distinguishing between near-field and far-field coupling effects. The near-field interaction, dominant at short distances, results from direct acoustic coupling between pipes, whereas the far-field coupling, relevant at larger separations, is mediated by radiated sound waves propagating through the surrounding medium. The interplay between these two mechanisms gives rise to nontrivial synchronization patterns that cannot be captured by models assuming instantaneous or purely local coupling. Nevertheless, for separation distances below 5 cm, significant discrepancies arise between the model and experiment, with the latter showing smaller synchronization plateaus. This suggests that the oscillator model requires refinement to accurately capture near-field interactions \cite{FIS16}. \\

To gain deeper theoretical insights, we employed two complementary analytical methods. The method of averaging yielded a generalized Adler equation describing the phase dynamics and equilibrium stability, enabling us to characterize the conditions for frequency locking. However, this approach did not determine the exact synchronization frequency. To address this, we applied the describing function method, which accurately predicted the synchronization frequency and elucidated the curvature of the frequency-detuning relationship observed in both numerical simulations and experiments. Together, these methods provided a comprehensive analytical framework for understanding both phase and frequency synchronization in the presence of delayed coupling.  \\

Our bifurcation analysis confirmed the existence of both in-phase and anti-phase synchronization, each associated with distinct synchronization frequencies in excellent agreement with theoretical predictions. The shape and stability of the Arnold tongue, which delineates the synchronization region in the parameter space of coupling strength and detuning, were found to be strongly influenced by the delay-induced interplay between near-field and far-field effects. The introduction of a delay-dependent coupling strength $\kappa(\tau)$ enabled us to capture key experimental observations, particularly the non-monotonic widening of the Arnold tongue as a function of $\tau$.  \\

The role of time delay in synchronization was found to be highly nontrivial. For small delays, the near-field coupling dominates, leading to a broader Arnold tongue, whereas at larger delays, far-field effects become more pronounced, resulting in sharper synchronization boundaries. Furthermore, at very small delays, an increased coupling strength enhances synchronization, leading to a pronounced widening of the Arnold tongue. At the same time, delay-dependent coupling significantly impacts the stability of different synchronization modes, notably rendering the in-phase mode unstable at zero delay. These findings underscore the necessity of incorporating both near- and far-field coupling effects when modeling acoustic interactions between organ pipes.  \\

In summary, our study highlights the fundamental role of time-delayed, spatially mediated coupling in shaping the synchronization of organ pipes. Despite the apparent simplicity of the Van der Pol oscillator model, it successfully captures the complex synchronization dynamics observed in experiments. By leveraging the Fast Fourier Transform, we extracted spectral features and validated numerical results against experimental data, demonstrating the model's predictive power in describing delay-induced nonlinear phenomena.  \\

Beyond organ pipes, our findings provide broader insights into coupled oscillatory systems where near- and far-field coupling mechanisms coexist. These results are relevant for various disciplines, including acoustics, physics, engineering, and biological systems, where delayed interactions play a crucial role in synchronization. Future work could refine the model further by incorporating additional aeroacoustic nonlinearities, distance-dependent coupling strengths, and experimental perturbations to deepen our understanding of synchronization transitions in spatially extended systems.

\section*{Acknowledgments}
We are grateful to Markus Abel, Jost Fischer, Markus Radke, Eckehard Sch{\"o}ll and Natalia Spitha for fruitful discussions. Special thanks go to the Hausorgel Arbeitskreis of the Gesellschaft der Orgelfreunde e. V., whose annual meeting in 2024 provided a forum for discussions on the results summarized here.


\end{document}